\documentclass[prd,twocolumn,preprintnumbers,nofootinbib,amsmath,amssymb,floatfix]{revtex4}
\usepackage{graphicx}
\usepackage{psfrag}
\usepackage{epstopdf}
\usepackage{verbatim}
\usepackage{rotating}
\usepackage{multirow}
\usepackage{hyperref}
\usepackage[letterpaper]{geometry}

\newcommand{\vesc}{v_\textrm{esc}}

\newcommand{\vq}{v_\textrm{q}}
\newcommand{\vlab}{v_\textrm{lab}}
\newcommand{\mchi}{m_\chi}

\newcommand{\mN}{m_\textrm{N}}
\newcommand{\sigmavH}{\sigma_\textrm{H}}
\newcommand{\dOq}{d\Omega_\textrm{q}}
\newcommand{\dOu}{d\Omega_\textrm{u}}
\newcommand{\dOn}{d\Omega_\textrm{n}}
\newcommand{\uhat}{\widehat{\mathbf{u}}}
\newcommand{\qhat}{\widehat{\mathbf{q}}}
\newcommand{\nhat}{\widehat{\mathbf{n}}}
\newcommand{\vhat}{\widehat{\mathbf{v}}}

\newcommand {\be} {\begin {equation}}
\newcommand {\ee} {\end {equation}}

\newcommand {\bes} {\begin {equation*}}
\newcommand {\ees} {\end {equation*}}

\newcommand{\es}[2] {\begin{equation} \label{#1} \begin{split} #2 \end{split} \end{equation}}

\begin{document}

\title{Harmonics in the Dark-Matter Sky:\\Directional Detection in the Fourier-Bessel Basis}
\author{Samuel K.\ Lee}
\affiliation{Princeton Center for Theoretical Science, Princeton University, Princeton, NJ 08544}
\email{samuelkl@princeton.edu}
\date{\today}

\begin{abstract}
Details about the velocity distribution of weakly interacting massive particle (WIMP) dark matter in our galaxy may be revealed by nuclear-recoil detectors with directional sensitivity.  Previous studies have assumed that the velocity distribution takes a simple functional form characterized by a small number of parameters.  More recent work has shown that basis-function expansions may allow for more general parameterization; such an approach has been considered for both the one-dimensional speed and momentum distributions, and also for three-dimensional velocity distributions obeying certain equilibrium conditions.  In this work, I extend this basis-function approach to allow for arbitrary velocity distributions by working in the Fourier-Bessel basis, deriving an analytic expression for the directional recoil spectrum.  Such an approach is completely general, and may be useful if the velocity distribution is too complex to be characterized by simple functional forms or is not completely virialized.  Results concerning the three-dimensional Radon transform of the Fourier-Bessel basis functions may be of general interest for tomographic applications.
\end{abstract}

\maketitle

\section{Introduction} \label{sec:intro}

The direct detection of WIMP dark matter by experiments sensitive to the energy spectrum of WIMP-induced nuclear recoils may be on the near horizon.  Although there have been several tantalizing hints of signals, a definitive detection has yet to be achieved.  Nevertheless, some theoretical and experimental efforts are already looking ahead to a new generation of direct-detection experiments, which employ detectors sensitive to the \emph{directional} recoil spectrum -- i.e., able to measure not only the energy but also the direction of nuclear recoils (for a review, see \cite{arXiv:0911.0323}).

A primary advantage of directional detectors is their ability to discriminate signal events from background events; the former should have a distinct directionality as a result of the WIMP ``wind'' induced by the sun's motion through the galaxy, whereas the latter should generally have an isotropic distribution \cite{Spergel:1987kx}.  The amplitude of the directional asymmetry can be of order unity, and is much larger than the few percent annual modulation in the signal that can be used by non-directional experiments to discriminate against background \cite{Drukier:1986tm}.  Yet despite this advantage, it is unlikely that directional experiments will be the first to detect and characterize the particle properties of dark matter, since their lower target densities and higher recoil-energy thresholds result in relatively smaller  event rates.  However, directional detectors will be in a unique position to probe the dark-matter velocity distribution in the post-discovery era.  They may allow for the study of ``WIMP astronomy,'' possibly revealing insights about structure formation on galactic scales via observations of the local dark-matter sky.  Furthermore, a detailed characterization of the dark-matter velocity distribution may be crucial for using direct-detection experiments to precisely constrain the dark-matter particle properties.

To this end, understanding how the WIMP velocity distribution affects the observed directional recoil spectrum is of primary importance.  Previous studies have been conducted along these lines, albeit with some simplifying assumptions.  For example, earlier work considered only velocity distributions with parameterized functional forms \cite{hep-ph/0209110,arXiv:0712.0053,arXiv:1110.0951,arXiv:1202.5035}.  The strength of this parametric, functional-form approach is perhaps the same as its weakness -- it reduces the amount of information to be constrained by directional experiments to only a few numbers, allowing for only simple models to be tested.

Recent studies on non-directional experiments have sought to move beyond the functional-form approach.  For example, Refs.~\cite{arXiv:1103.5145,arXiv:1207.2039} explored the possibility of constraining either the one-dimensional speed or momentum distribution in a series of bins, assuming that the distribution is constant in each bin.  An alternative approach is to perform a basis-function expansion of the speed distribution, which is then completely characterized by its expansion coefficients \cite{arXiv:1303.6868,arXiv:1310.7039,arXiv:1312.1852}.

This latter approach can also be applied to the full three-dimensional velocity distribution in studies of directional experiments.  In this context, Ref.~\cite{arXiv:1204.5487} examined a basis consisting of functions of the integrals of motion $E$, $L$, and $L_z$.  Such a basis is convenient for two reasons: 1) if the velocity distribution is at equilibrium (virialized) and can hence be written as a function of integrals of motion, measuring the dependence of the local distribution on these integrals of motion allows us to infer the global distribution (throughout the galaxy), and 2) if the velocity distribution is separable in $E$, $L$, and $L_z$, the basis functions can be chosen so that the resulting directional recoil spectrum can be found analytically.  However, we see that we require the assumptions of equilibrium and separability in order to make use of this integrals-of-motion basis.  

One motivation of such basis-function approaches has been to remove the biases incurred by the choice of the form of the velocity distribution in the functional-form approach, moving towards a more agnostic description of the velocity distribution.  As such, the natural continuation is to consider a basis that allows for completely general velocity distributions.  In this paper, we explore an expansion of the velocity distribution in the Fourier-Bessel basis.  The Fourier-Bessel basis indeed allows for complete generality, so that we can drop the assumptions of equilibrium and separability required by Ref.~\cite{arXiv:1204.5487}.  Furthermore, we shall see that we can still achieve an analytic expression for the directional recoil spectrum.

In the next Section, we will briefly review basic expressions for the directional recoil spectrum of WIMP-induced nuclear recoils.  In Sec.~\ref{sec:FB}, we will discuss the Fourier-Bessel basis and its mathematical properties.  Sec.~\ref{sec:DDinFB} will elucidate the application of the Fourier-Bessel expansion to directional detection, and will derive an analytic expression for the predicted directional recoil spectrum in terms of the velocity-distribution Fourier-Bessel coefficients.  Appendices defining mathematical notation and addressing some additional points follow the concluding discussion in Sec.~\ref{sec:conclusions}.

\section{Directional Detection: Basics} \label{sec:basics}

We are ultimately interested in determining the Galactic-frame velocity distribution function by a measurement of the directional recoil spectrum of WIMP-induced nuclear recoils.  Let us first examine how these two quantities are related.

The Galactic-frame velocities of the WIMP and the lab are related to the lab-frame WIMP velocity $\mathbf{v}$ by a simple Galilean transformation, such that the Galactic-frame WIMP velocity is $\mathbf{v} + \mathbf{\vlab}$.  Hence, the Galactic-frame velocity distribution $g$ and the lab-frame velocity distribution $f$ are simply related by a translation,
\es{eq:f-g}{
f(\mathbf{v}) = g(\mathbf{v}+\mathbf{\vlab})\,.
}

Assuming elastic WIMP-nucleus scattering, Ref.~\cite{hep-ph/0209110} showed that the directional recoil spectrum is given by
\es{eq:spectrum}{
\frac{dR}{dE \dOq} = \frac{\rho_0 \sigma_\mathrm{N} S(q)}{4\pi \mchi \mu_\mathrm{N}^2} \widehat{f}(\vq, \qhat)\,.
}
Here, $R$ is the number of events per exposure (detector mass multiplied by time), $\rho_0$ is the local WIMP density, ${\mu_\mathrm{N} = \mchi \mN / (\mchi + \mN)}$ is the reduced mass of the WIMP-nucleus system, ${\mathbf{q}=q\qhat}$ is the lab-frame nuclear-recoil momentum, ${E = q^2/2\mN}$ is the lab-frame nuclear-recoil energy, and ${\vq = q/2\mu_\mathrm{N}}$ is the minimum lab-frame WIMP speed required to yield a recoil energy $E$.\footnote{Inelastic scattering is easily accounted for by changing the relation between $\vq$ and $E$ appropriately.}  Furthermore, we have written the WIMP-nucleus elastic cross section as ${d\sigma/dq^2 = \sigma_\mathrm{N} S(q) / 4 \mu_\mathrm{N}^2 v^2}$ (where $S(q)$ is the nuclear form factor), and 
\es{eq:radon}{
 \widehat{f}(\vq, \qhat) &= \int_\mathcal{P}\! d\mathcal{P}\, f(\mathbf{v})\\ &= \int\! d^3v\, \delta(\vq - \mathbf{v}\cdot\qhat) f(\mathbf{v})
}
is the Radon transform of the lab-frame WIMP velocity distribution.\footnote{The Radon transform is widely used in tomographic applications involving imaging by sections (planes, in the case of three dimensions); e.g., magnetic resonance imaging.  For an overview, see Ref.~\cite{deans2007radon}.}  These relations follow from the kinematics; the rate of nuclear recoils with a given energy $E$ and direction $\qhat$ is proportional to the number of WIMPs with velocities $\mathbf{v}$ that can induce such recoils  -- i.e., that satisfy the non-relativistic elastic-scattering kinematics determined by $\vq$. Such velocities lie on the plane $\mathcal{P}$ in the lab-frame velocity space that is defined by ${\mathbf{v}\cdot\qhat = \vq}$, and are selected by the delta function in the Radon transform.\footnote{In general, the presence of this delta function makes difficult the calculation of the directional recoil spectrum for arbitrary velocity distributions via numerical integration of Eq.~\eqref{eq:radon}.  We shall see that use of the Fourier-Bessel basis allows for an analytic result.}

From the perspective of particle physics, the primary quantities of interest to be determined by direct-detection experiments are the WIMP particle properties $\mchi$ and $\sigma_\mathrm{N}$.  However, from the perspective of astrophysics, the determination of the velocity distribution $g$ is also of interest; for example, galactic-scale N-body simulations may give predictions for $g$.  Furthermore, an understanding of $g$ may be critical in correctly interpreting the results of direct-detection experiments and accurately constraining the WIMP particle properties.  In what follows, our primary goal will be to examine what can be learned about $g$ from the directional recoil spectrum $dR/dE \dOq$.

\section{Fourier-Bessel Decomposition of the Velocity Distribution} \label{sec:FB}

If our goal is to determine the velocity distribution $g$, we must first understand how it may be characterized.  Most direct-detection studies assume that $g$ can be approximated by a truncated Maxwellian; such a velocity distribution is isotropic and is appropriate for collisionless particles with a halo density profile that falls as ${\rho \propto r^{-2}}$.  This ``Standard Halo Model'' velocity distribution may be specified by only two parameters: the velocity dispersion $\sigmavH$ and the truncating escape velocity $\vesc$.  Likewise, more complex velocity distributions -- which may be non-Maxwellian \cite{arXiv:1010.4300,arXiv:1304.6401}, anisotropic \cite{astro-ph/0209528,arXiv:0704.2909,arXiv:0812.0362,arXiv:1311.5477}, or include the presence of a dark-matter disk \cite{arXiv:0803.2714,arXiv:0804.2896,arXiv:0902.0009,arXiv:1207.1050,arXiv:1303.1521,arXiv:1303.3271,arXiv:1307.4095,arXiv:1308.1703}, streams \cite{astro-ph/0309279,astro-ph/0607121,arXiv:1109.0014,arXiv:1203.6617}, debris flows \cite{arXiv:1105.4166,arXiv:1202.0007}, or extragalactic components \cite{astro-ph/0106480,arXiv:1208.0392} -- may be modeled by using similarly manageable functional forms with a few additional parameters.  These parameters may then be constrained by measurement of the directional recoil spectrum.

However, it is clear that such studies are dependent on the \textit{a priori} functional form chosen for $g$, which may be oversimplified.  It might then be advantageous to instead consider an alternative approach in which we use a complete basis for functions of velocity to decompose $g$.  We can then study completely general distribution functions.

\subsection{The Fourier-Bessel Basis} \label{sec:FBbasis}
In this paper, we will show that a convenient basis is given by a generalization of the Fourier-Bessel basis.  We can define basis functions for functions of velocity $g(\mathbf{v})$ simply given by the product of a spherical Bessel function and a real spherical harmonic
\es{eq:FBfunction}{
\Psi^u_{lm}(\mathbf{v}) = 4\pi j_l(uv) S_{lm}(\vhat)\,,
}
such that the basis functions are labeled by a real number $u$ and integers $l$ and $m$.  For the purposes of presentation, the real spherical harmonics $S_{lm}$ (defined in Appendix~\ref{app:realSH}) are used instead of the usual spherical harmonics $Y_{lm}$, since we are primarily concerned with expanding the real velocity distribution.

The functions $\Psi^u_{lm}(\mathbf{v})$ are solutions to the scalar Helmholtz equation ${(\nabla^2 + u^2)\Psi^u_{lm}(\mathbf{v}) = 0}$, and obey the completeness relation
\es{}{
\sum_{lm} \int\! \frac{u^2 du}{(2\pi)^3} \Psi^u_{lm}(\mathbf{v})  \Psi^u_{lm}(\mathbf{v}') = \delta^3(\mathbf{v}-\mathbf{v}')
}
and the orthonormality relation
\es{eq:ortho}{
\int\! d^3v\,  \Psi^u_{lm}(\mathbf{v}) \Psi^{u'}_{l'm'}(\mathbf{v}) = \frac{(2\pi)^3}{u^2} \delta(u - u') \delta_{ll'} \delta_{mm'}\,.
}

The $\Psi^u_{lm}(\mathbf{v})$ thus constitute a complete orthonormal basis for velocity distributions, any of which can be decomposed as
\es{}{
g(\mathbf{v}) = \sum_{lm} \int\! \frac{u^2 du}{(2\pi)^3} \Psi^u_{lm}(\mathbf{v}) g^u_{lm} \,.
}
We may also invert this relation to obtain an expression for the expansion coefficients,
\es{eq:FB-coeff}{
g^u_{lm} = \int\! d^3v\,  \Psi^u_{lm}(\mathbf{v}) g(\mathbf{v}) \,.
}

\subsection{Relation to Fourier Basis} \label{sec:fourierbasis}
Alternatively, one might use the standard Fourier basis to decompose the velocity distribution into plane waves,
\es{eq:F-exp}{
g(\mathbf{v}) = \int\! \frac{d^3u}{(2\pi)^3}\, e^{i \mathbf{u}\cdot\mathbf{v}} \widetilde{g}(\mathbf{u}) \,,
}
with expansion coefficients
\es{}{
\widetilde{g}(\mathbf{u}) = \int\! d^3v\, e^{-i \mathbf{u}\cdot\mathbf{v}} g(\mathbf{v}) \,.
}
For our purposes, this is of special interest because of the Fourier slice theorem that relates the Fourier transform $\widetilde{g}$ to the Radon transform $\widehat{g}$ via
\es{eq:Fourier-slice}{
\widetilde{g}(\mathbf{u}) = \int\! d\vq\, e^{-i u \vq} \widehat{g}(\vq, \uhat) \,.
}
Note that the integral includes negative values of $\vq$ here.

Furthermore, examining ~Eqs.~\eqref{eq:FB-coeff}~and~\eqref{eq:F-exp}, taking note of the plane-wave expansion
\es{eq:pwe}{
e^{i \mathbf{u}\cdot\mathbf{v}} &= \sum_{lm} 4\pi i^l j_l(uv) S_{lm}(\uhat) S_{lm}(\vhat)\\
		&= \sum_{lm} i^l \Psi^u_{lm}(\mathbf{v}) S_{lm}(\uhat)\,,
}
and using the orthonormality relation of Eq.~\eqref{eq:ortho}, it can be shown that the Fourier-Bessel and Fourier coefficients are simply related by
\es{eq:FB-F-coeff}{
g^u_{lm} = i^l \int\! \dOu\, S_{lm}(\uhat) \widetilde{g}(\mathbf{u}) \,.
}
That is, the Fourier-Bessel coefficient is given by multiplying the Fourier coefficient by the corresponding spherical harmonic and performing the angular integral.

\subsection{Boundary Conditions} \label{sec:BC}

The Galactic-frame WIMP velocity distribution is typically truncated at the escape velocity $\vesc$.  Thus, we now consider the decomposition for the case in which the function $g(\mathbf{v})$ vanishes for ${v > \vesc}$ \cite{astro-ph/9406009}.  This boundary condition constrains the allowed radial eigenfunctions and generates a discrete radial spectrum with eigenvalues $\{u_{ln}\}$ that satisfy
\es{}{
j_l\left(u_{ln} \vesc\right) = 0\,.
}
We see that ${u_{ln} = x_{ln} / \vesc}$, where $x_{ln}$ is the $n$th zero of $j_l$.

Functions that obey this boundary condition can be expanded using the basis functions
\es{eq:FBfunctiondisc}{
\Psi^n_{lm}(\mathbf{v}) &= 4\pi c_{ln} j_l (u_{ln} v) S_{lm}(\vhat) \theta(\vesc-v)\\
&= c_{ln} \Psi^{u_{ln}}_{lm}(\mathbf{v})\theta(\vesc-v)\,,
}
which are labeled by integers $n$, $l$, and $m$.  Here,
\es{}{
c^{-1}_{ln} \equiv \frac{x_{ln} j_{l+1}(x_{ln})}{\sqrt{\pi}}
}
is a normalization factor, such that these functions obey the completeness relation
\es{}{
\sum_{nlm} \frac{u_{ln}^2 \vesc^{-1}}{(2\pi)^3} \Psi^n_{lm}(\mathbf{v}) \Psi^n_{lm}(\mathbf{v}') = \delta^3(\mathbf{v}-\mathbf{v}')
}
and the orthonormality relation
\es{}{
\int\! d^3v\, \Psi^n_{lm}(\mathbf{v}) \Psi^{n'}_{l'm'}(\mathbf{v}) = \frac{(2\pi)^3}{u_{ln}^2 \vesc^{-1}} \delta_{nn'} \delta_{ll'} \delta_{mm'}\,.
}

A velocity distribution truncated at $\vesc$ can therefore be expanded as
\es{eq:g-exp}{
g(\mathbf{v}) = \sum_{nlm} \frac{u_{ln}^2 \vesc^{-1}}{(2\pi)^3} \Psi^n_{lm}(\mathbf{v}) g^n_{lm}\,,
}
with the inverse relation giving the discrete expansion coefficients
\es{eq:discCoeff}{
g^n_{lm} &= \int\! d^3v\, \Psi^n_{lm}(\mathbf{v}) g(\mathbf{v}) \\
&= c_{ln} g^{u_{ln}}_{lm}\,.
}
Comparing with the expressions in Sec.~\ref{sec:FBbasis}, it is easy to see that the discrete and continuous expressions are related by ${\Psi^n_{lm} \to \Psi^u_{lm}}$, ${u_{ln} \to u}$, and ${\sum_n \vesc^{-1} \to \int\! du}$ in the limit that ${\vesc \to \infty}$.  Working with the discrete expressions has the obvious advantage of limiting the expansion coefficients $g^n_{lm}$ to a countable set, which can be truncated to give a finite number of quantities to be determined.

\subsection{Advantages and Disadvantages} \label{sec:FBadv}

One clear advantage of the Fourier-Bessel basis is that it allows the representation of general functions of velocity.  By moving beyond \emph{a priori} assumptions of simple functional forms, we can avoid biasing inferences drawn from direct-detection experiments.  Furthermore, unlike with the integrals-of-motion basis used in Ref.~\cite{arXiv:1204.5487}, we are not limited to equilibrium velocity distributions that are separable in the integrals of motion.  Although it is true that the assumption of equilibrium allows one to infer the global velocity distribution from measurements of the local velocity distribution, it is not clear if this assumption is valid on the scales probed by direct-detection experiments.  The generality of the Fourier-Bessel basis allows us to remain completely agnostic about the form of the velocity distribution.

Our ultimate goal is then to recover the discrete Fourier-Bessel coefficients $g^n_{lm}$ from a directional experiment.  In an ideal experiment with nearly continuous data and complete spectral coverage, one could use traditional tomographic methods to simply invert the Radon transform and deduce the underlying $g^n_{lm}$ (see Appendix~\ref{app:invert}).  Realistically, we are limited to using maximum-likelihood methods to scan over the parameter space of the coefficients $g^n_{lm}$, searching for those coefficients that yield the best-fit predicted directional recoil spectrum for a given data set; similar analyses were carried out for simulated data sets for both non-directional and directional experiments in Refs.~\cite{arXiv:1204.5487,arXiv:1303.6868,arXiv:1310.7039,arXiv:1312.1852}.  Since we must limit the parameter space according to the available computational power by truncating at some $n$ and $l$, it is possible that only large-scale features in $g$ may be reconstructed at first.  This is an unfortunate drawback of the basis-function and maximum-likelihood approach.  Nevertheless, in order to construct the likelihood function, one only needs to know how to calculate the predicted directional recoil spectrum given the coefficients $g^n_{lm}$, and can simply reconstruct $g$ from the maximum-likelihood coefficients using Eq.~\eqref{eq:g-exp}.

As we shall demonstrate shortly, the Fourier slice theorem in Eq.~\eqref{eq:Fourier-slice} and the unique appearance of the Fourier-Bessel basis functions in the plane-wave expansion in Eq.~\eqref{eq:pwe} ultimately allow us to find an analytic expression for the Radon transform of the basis functions.  The upshot of this is that given the expansion coefficients $g^n_{lm}$ for an arbitrary velocity distribution $g$, there is a simple analytic expression that yields the predicted directional recoil spectrum, allowing us to easily construct the desired likelihood function.  This avoids the need for computing the Radon transform numerically, a process made ungainly by the presence of the delta function in Eq.~\eqref{eq:radon}.  This is the primary reason the Fourier-Bessel basis might be a unique and useful basis for directional-detection analyses.

However, it could be argued that numerical integration is still required to find the expansion coefficients $g^n_{lm}$ for a given distribution $g$ via Eq.~\eqref{eq:discCoeff}.  For example, one may be interested in calculating the coefficients $g^n_{lm}$ corresponding to the distribution $g$ predicted by an N-body simulation (in order to compare with those coefficients recovered from a directional experiment, perhaps).  Fortunately, the Fourier-Bessel basis is well studied, so that efficient routines to calculate Fourier-Bessel coefficients are readily available (e.g., see Refs.~\cite{arXiv:1111.3591,arXiv:1112.0561}).  Alternatively, recent work has shown that the Fourier-Bessel expansion can be simply related to an analogous Fourier-Laguerre expansion.  Fortunately, there exists an exact quadrature rule for the evaluation of the integrals for the Fourier-Laguerre coefficients (i.e., integrals analogous to Eq.~\eqref{eq:discCoeff}), and a corresponding sampling theorem can be found \cite{arXiv:1205.0792,arXiv:1308.5480}.  Thus, if the velocity distribution is band-limited in the Fourier-Laguerre basis, one can easily find the Fourier-Bessel coefficients using these results.  In any case, such concerns are secondary if we are primarily concerned with computing the directional recoil spectrum, which we shall now do.

\section{The Directional Recoil Spectrum in the Fourier-Bessel Basis} \label{sec:DDinFB}

Our goal is to calculate the predicted directional recoil spectrum $dR/dE \dOq$ given an arbitrary Galactic-frame dark-matter velocity distribution $g$ truncated at $\vesc$.  We shall see that utilization of the Fourier-Bessel basis results in a straightforward expression for $dR/dE \dOq$ in terms of the expansion coefficients $g^n_{lm}$.

First, because the lab-frame distribution $f$ and the Galactic-frame distribution $g$ are related by a translation as in Eq.~\eqref{eq:f-g}, the behavior of the Radon transform under translation implies that
\es{eq:Rf-Rg}{
\widehat{f}(\vq, \uhat) = \widehat{g}(\vq + \mathbf{\vlab}\cdot\uhat,\uhat)\,.
}
Thus, for a given $\mathbf{\vlab}$, we can easily find $\widehat{f}$ once we have first calculated $\widehat{g}$.  Noting that the Radon transform is linear, we can expand $g$ as in Eq.~\eqref{eq:g-exp} and find 
\es{eq:ghat-exp}{
\widehat{g}(\vq, \uhat) = \sum_{nlm} \frac{u_{ln}^2 \vesc^{-1}}{(2\pi)^3} \widehat{\Psi}^n_{lm}(\vq, \uhat) g^n_{lm}\,.
}
We then see that the problem reduces to calculating the Radon transform $\widehat{\Psi}^n_{lm}$ of the Fourier-Bessel basis functions.  However, recall that the Fourier slice theorem relates the Radon transform $\widehat{\Psi}^n_{lm}$ to the Fourier transform $\widetilde{\Psi}^n_{lm}$ via Eq.~\eqref{eq:Fourier-slice}, the inverse of which yields
\es{eq:Psi-Fourier-slice-inverse}{
\widehat{\Psi}^n_{lm}(\vq, \uhat) = \frac{1}{2\pi} \int\! du\, e^{i u \vq} \widetilde{\Psi}^n_{lm}(\mathbf{u})\,.
}

\begin{figure*}[t]
\includegraphics[trim=0 0 0 0,clip,width=\textwidth]{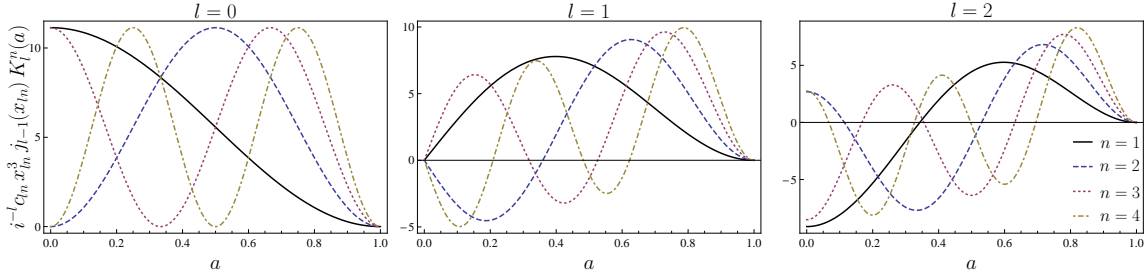}
\caption{The functions $K^n_l(a)$ (normalized by the $n$- and $l$-dependent factors present in Eq.~\eqref{eq:key-result}) are plotted for a few values of $n$ and $l$.  Note that the $K^n_l$ are simply sinusoidal for $l=0$ and are combinations of sinusoidal and polynomial terms for higher $l$.  The $\vq$ dependence of the directional recoil spectrum is given by a linear combination of these functions evaluated at ${a = (\vq + \mathbf{\vlab}\cdot\qhat)/\vesc}$.}
\label{fig:Knl}
\end{figure*}

Thus, the crux of the problem of finding $\widehat{f}$ ultimately reduces to first calculating the three-dimensional Fourier transform $\widetilde{\Psi}^n_{lm}$ and then taking an additional one-dimensional Fourier transform to yield $\widehat{\Psi}^n_{lm}$.  The first step is accomplished by using the plane-wave expansion of Eq.~\eqref{eq:pwe},
\es{eq:PsiFT}{
\widetilde{\Psi}^n_{lm}(\mathbf{u}) &= \int\! d^3v\, e^{-i \mathbf{u}\cdot\mathbf{v}} \Psi^n_{lm}(\mathbf{v})\\
	&= 16 \pi^2 i^{-l} c_{ln} j_{l-1} (x_{ln})\\ &\indent\indent\times \frac{u_{ln} \vesc^2}{u^2-u_{ln}^2} j_l(u \vesc) S_{lm}(\uhat)\,,
}
where we have have used a result for the integral ${\int_0^{\vesc}\! dv\,  v^2 j(u_{ln} v) j_l(u v)}$ and the orthonormality of the $S_{lm}$ to simplify.  The second step, accomplished by substituting Eq.~\eqref{eq:PsiFT} into Eq.~\eqref{eq:Psi-Fourier-slice-inverse}, then yields
\es{eq:PsiRT}{
\widehat{\Psi}^n_{lm}(\vq, \uhat) &= 8\pi \vesc^2 i^{-l} c_{ln} x_{ln} j_{l-1}(x_{ln})\\&\indent\indent\times K^n_l(\vq/\vesc) S_{lm}(\uhat)\,,
}
where we have defined the integrals
\begin{widetext}
\es{eq:Knl}{
K^n_l(a) &= \int\! dx\, \frac{e^{i a x} j_l(x)}{x^2-x_{ln}^2}\\
	&= \frac{1}{2} \sum_{r=1}^{l+1} \sum_{s=\pm 1} \left[(\alpha_{lr} - i s \beta_{lr}) \operatorname{Re} I(a+s,r,x_{ln}) +(i \alpha_{lr} + s \beta_{lr}) \operatorname{Im} I(a+s,r,x_{ln}) \right]\,.
}
\end{widetext}
The calculation of the integrals $K^n_l(a)$ requires contour integration and is somewhat involved, so we relegate the derivation of this result to Appendix~\ref{app:contourint}.  The coefficients $\alpha_{lr}$ and $\beta_{lr}$ are also defined there, as is the function ${I(a+s,r,x_{ln})}$.  However, we will mention here that $\alpha_{lr}$, $\beta_{lr}$, and ${I(a+s,r,x_{ln})}$ are such that $i^{-l} K^n_l$ is always real, so that $\widehat{\Psi}^n_{lm}$ is also real, as expected.  The functions $K^n_l$ are plotted for a few values of $n$ and $l$ in Fig.~\ref{fig:Knl}.

Although these equations may appear relatively complicated, they are actually surprisingly straightforward.  The dependence of $\widehat{\Psi}^n_{lm}(\vq, \uhat)$ on $\vq$ and $\uhat$ is wholly contained in the functions $K^n_l(\vq/\vesc)$ and $S_{lm}(\uhat)$, respectively; the other factors appearing in Eq.~\eqref{eq:PsiRT} simply normalize the function and are easily calculable.   It is difficult to think of a more generic and convenient angular dependence than that given by the spherical harmonics.  Furthermore, as shown in Appendix~\ref{app:contourint}, $K^n_l$ is actually composed of a finite sum of elementary functions, so the dependence on $\vq$ can be easily calculated.  Again, that we are able to achieve a relatively simple analytic result for the Radon transform of the Fourier-Bessel basis functions is partly due to the unique appearance of these basis functions in the plane-wave expansion, as advertised in Sec.~\ref{sec:FBadv}.

The key result of this paper is then given by combining Eqs.~\eqref{eq:spectrum}, \eqref{eq:Rf-Rg}, \eqref{eq:ghat-exp}, \eqref{eq:PsiRT}, and \eqref{eq:Knl},
\begin{widetext}
\es{eq:key-result}{
\frac{dR}{dE \dOq} = \frac{\rho_0 \sigma_\mathrm{N} S(q)}{4\pi^3 \mchi \mu_\mathrm{N}^2 \vesc} \sum_{nlm} i^{-l} c_{ln} x_{ln}^3 j_{l-1}(x_{ln}) K^n_l[(\vq + \mathbf{\vlab}\cdot\qhat)/\vesc] S_{lm}(\qhat) g^n_{lm}\,.
}
\end{widetext}
This expression gives the directional recoil spectrum $dR/dE \dOq$ expected from an arbitrary Galactic-frame velocity distribution $g$ characterized by the Fourier-Bessel expansion coefficients $g^n_{lm}$.  Given the coefficients $g^n_{lm}$, we see that the energy dependence of the directional recoil spectrum is then determined by the functions $K^n_l$ and the nuclear form factor $S(q)$.

Alternatively, one may instead choose to write Eq.~\eqref{eq:key-result} in terms of the expansion coefficients $f^n_{lm}$ for the lab-frame velocity distribution $f$.  Besides the replacement of $g^n_{lm}$ with $f^n_{lm}$, the rest of the expression is similar except for two differences: 1) basis functions truncated at ${\vesc' = \vesc + \vlab}$ instead of at $\vesc$ should be used throughout, and 2) $K^n_l(\vq/\vesc')$ appears instead of ${K^n_l[(\vq + \mathbf{\vlab}\cdot\qhat)/\vesc]}$.  The second difference is especially noteworthy, since the angular dependence of $dR/dE\dOq$ is then completely contained in the real spherical harmonics $S_{lm}(\qhat)$.  This suggests that a spherical-harmonic decomposition of the directional recoil spectrum might be useful, since the moments $(dR/dE\dOq)_{lm}$ are then generated only by the coefficients $f^n_{lm}$ with the same $l$ and $m$ indices.  One might then try to choose a coordinate system in which some of either the coefficients $f^n_{lm}$ or the observed moments $(dR/dE\dOq)_{lm}$ vanish or are relatively small.  For example, if the distribution $f$ is symmetric about the axis defined by $\mathbf{\vlab}$, then $f$ contains only zonal harmonics in the coordinate system with $\widehat{\mathbf{z}} || \mathbf{\vlab}$, in which case the coefficients and observed moments with ${m \neq 0}$ vanish.
 
\section{Discussion} \label{sec:conclusions}

Directional detectors will play a key role in the post-discovery era of WIMP direct-detection experiments.  They may allow for the mapping of the local WIMP velocity distribution, which may yield both insights on galactic structure formation and improved constraints on the WIMP particle properties.

We have shown that a convenient parameterization of the velocity distribution is provided by an expansion in the familiar Fourier-Bessel basis.  Unlike other bases explored in previous work, the Fourier-Bessel basis allows for the representation of general velocity distributions, without requiring assumptions of equilibrium or separability.  More importantly, an analytic expression for the predicted directed recoil spectrum in terms of the Fourier-Bessel coefficients of the velocity distribution can be derived, which allows for the calculation of event rates without requiring numerical integration of the Radon transform.

As previously mentioned, one way to explore the power of this formalism is to consider a fiducial dark-matter velocity distribution, simulate nuclear-recoil events and the corresponding directional recoil spectrum, and then attempt to recover the Fourier-Bessel coefficients of the distribution via maximum-likelihood methods.  One might consider fiducial distributions motivated by the results of galactic-scale N-body simulations.  We leave such studies to future work.

Furthermore, the harmonic decomposition of the angular and energy dependence of the event rate we have studied here might also be combined with recent work studying the harmonic decomposition of the time dependence of the rate \cite{arXiv:1111.4222,arXiv:1307.5323}.  This is simply accomplished within the framework we have constructed by allowing $\mathbf{\vlab}$ to be a time-dependent quantity.  It may then be important to also account for the gravitational focusing of WIMPs by the sun, which affects all three of these dependences \cite{arXiv:1308.1953}.

In practice, it is likely that analyses of data in the post-discovery era will begin by assuming that the velocity distribution takes a simple functional form, progressing to more general basis-function methods only after the basic features of the velocity distribution have been discerned.  It may then be some time before the power of the integrals-of-motion and Fourier-Bessel bases, or other yet unexplored bases, can be fully utilized in directional analyses.  Nevertheless, with this paper we have prepared the necessary framework for such future studies, which may ultimately be crucial in revealing the nature of the dark matter on scales both microscopic and galactic.

\acknowledgments

This work was initially inspired by Ref.~\cite{arXiv:1209.0761}, which explored the use of the Fourier-Bessel functions and their higher-spin analogues in cosmology; a prepublication draft of that paper was kindly provided by Liang Dai, Donghui Jeong, and Marc Kamionkowski, as were helpful comments. The author also thanks Adam Burrows for pointing out Ref.~\cite{arXiv:1111.3591}, as well as Mariangela Lisanti and Annika Peter for useful remarks on various drafts of this paper.

\appendix
\section{Real Spherical Harmonics} \label{app:realSH}

Since the velocity distribution function $g(\mathbf{v})$ is real, we choose our Fourier-Bessel eigenfunctions in Eqs.~\eqref{eq:FBfunction}~and~\eqref{eq:FBfunctiondisc} to contain \emph{real} spherical harmonics $S_{lm}$ (as opposed to the usual spherical harmonics $Y_{lm}$).  This results in real expansion coefficients $g^u_{lm}$ and $g^n_{lm}$.

We define our real spherical harmonics to be
\es{}{
S_{lm}(\nhat) &= 
\begin{cases}
\sqrt{2} N_{lm} P_l^m(\cos \theta) \cos m\phi & m > 0\\
N_{l0} P_l(\cos \theta) & m = 0\\
\sqrt{2} N_{lm} P_l^{m}(\cos \theta) \sin m\phi & m < 0\\
\end{cases}\\
&= 
\begin{cases}
\frac{1}{\sqrt{2}}\left[Y_{lm}(\nhat) + Y^*_{lm}(\nhat)\right] & m > 0\\
Y_{l0}(\nhat) & m = 0\\
-\frac{i}{\sqrt{2}}\left[Y_{lm}(\nhat) - Y^*_{lm}(\nhat)\right] & m < 0\\
\end{cases}\\
& = A_{m} \left[Y_{lm}(\nhat) + B_m Y^*_{lm}(\nhat)\right]  \,,
}
where $P_l^m$ are associated Legendre polynomials,
\es{}{
 N_{lm} = \sqrt{\frac{(2l+1)}{4\pi}\frac{(l-m)!}{(l+m)!}}
}
is the usual normalization factor (such that the $S_{lm}$ obey the orthonormality relation ${\int\! \dOn\, S_{lm}(\nhat) S_{l'm'}(\nhat) = \delta_{ll'} \delta_{mm'}}$), and
\es{}{
A_{m} &= 
\begin{cases}
1/\sqrt{2} & m > 0\\
1/2 & m = 0\\
-i/\sqrt{2} & m < 0\\
\end{cases}\\
&= B_m A^*_m
}
and
\es{}{
B_{m} = 
\begin{cases}
1 & m \geq 0\\
-1 & m < 0\\
\end{cases}
}
are constants we define here to condense the notation that follows.

The $S_{lm}$ provide a complete, orthonormal basis for square-integrable real functions on the sphere.  Such a function $f(\nhat)$ can then be decomposed as
\es{}{
f(\nhat) = \sum_{lm} \alpha_{lm} S_{lm}(\nhat)\,,
}
where the real-spherical-harmonic expansion coefficients ${\alpha_{lm} = \int\! \dOn\, S_{lm}(\nhat) f(\nhat)}$ are related to the usual spherical-harmonic expansion coefficients ${a_{lm} = \int\! \dOn\, Y^*_{lm}(\nhat) f(\nhat)}$ by ${\alpha_{lm} = A_m a^*_{lm} + A^*_m a_{lm}}$.

Consider the Gaunt coefficient given by the angular integral of the product of three spherical harmonics, which may be written in terms of Wigner-3$j$ symbols
\es{}{
G(L L' L'') &= \int\! \dOu\, Y_{lm}(\uhat) Y_{l'm'}(\uhat) Y_{l''m''}(\uhat)\\
  &= \frac{[l][l'][l'']}{\sqrt{4\pi}} \begin{pmatrix} l & l' & l'' \\ 0 & 0 & 0 \end{pmatrix} \begin{pmatrix} l & l' & l'' \\ m & m' & m'' \end{pmatrix}\,, 
}
where ${[l] \equiv \sqrt{2l+1}}$ and ${L \equiv (l, m)}$.  We may define an analogous coefficient for the integral of three real spherical harmonics,
\es{}{
H(L L' L'') &= \int\! \dOu\, S_{lm}(\uhat) S_{l'm'}(\uhat) S_{l''m''}(\uhat)\\
&=  A_m A_{m'} A_{m''} (1 + B_m B_{m'} B_{m''})\\
&\indent\indent\times [ G(L L' L'') \\
&\indent\indent\indent+ (-1)^m B_m G(\bar{L} L' L'') \\
&\indent\indent\indent+ (-1)^{m'} B_{m'} G(L \bar{L'} L'')\\
&\indent\indent\indent+ (-1)^{m''} B_{m''} G(L L' \bar{L''})]\,,
}
where we have introduced ${\bar{L} \equiv (l,-m)}$ and used the selection rules governing the Gaunt coefficients to simplify the final expression.  Note that $H(L L' L'')$ is non-zero only for ${m + m' + m''}$ even.

\section{Contour Integration} \label{app:contourint}

Our goal here will be to calculate the integral
\es{eq:Knl-app}{
K^n_l(a) = \int\! dx\, \frac{e^{i a x} j_l(x)}{x^2-x_{ln}^2}
}
that appears in Eq.~\eqref{eq:PsiRT} and determines the $\vq$ dependence of ${\widehat{\Psi}^n_{lm}(\vq, \uhat) \propto K^n_l(\vq/\vesc)}$.  We shall follow the methods used in Ref.~\cite{Haynes1997}, which investigated the use of the Fourier-Bessel basis in modeling pseudopotentials for calculations in density function theory.

We start by expanding the spherical Bessel function appearing in the numerator of the integrand,
\es{}{
j_l(x) =  \sum_{r=1}^{l+1}  \frac{\alpha_{lr} \cos x  + \beta_{lr} \sin x}{x^r} \,,
}
where the expansion coefficients $\alpha_{lr}$ and $\beta_{lr}$ can be found by using the standard recurrence relation
${j_l(x) = \frac{2l-1}{x} j_{l-1}(x) - j_{l-2}(x)}$, with ${j_0(x) = \frac{\sin x}{x}}$ and ${j_1(x) = \frac{\sin x}{x^2} - \frac{\cos x}{x}}$.

Writing the exponential appearing in the numerator of the integrand of Eq.~\eqref{eq:Knl-app} as ${e^{i a x} = \cos ax + i \sin ax}$, we see that the integral becomes
\begin{widetext}
\es{eq:Knl-sum}{
K^n_l(a) &= \sum_{r=1}^{l+1} \int\! dx\, \frac{\alpha_{lr} \cos ax \cos x + \beta_{lr} \cos ax \sin x + i \left( \alpha_{lr} \sin ax \cos x + \beta_{lr} \sin ax \sin x \right) }{x^r(x^2-x_{ln}^2)} \\
	&= \frac{1}{2} \sum_{r=1}^{l+1} \sum_{s = \pm 1} \int\! dx\, \frac{(\alpha_{lr} - i s \beta_{lr}) \cos [(a + s)x] + (i \alpha_{lr} + s \beta_{lr}) \sin [(a + s)x]}{x^r (x^2-x_{ln}^2)}\,.
}
\end{widetext}
Here, we have used trigonometric addition identities to simplify the numerator of the integrand.

Examining Eq.~\eqref{eq:Knl-sum}, we see that the integral $K^n_l$ is itself a sum of terms proportional to the real and imaginary parts of integrals of the form
\es{eq:singular}{
\int\! dx\, \frac{e^{i A x}}{x^r (x^2-x_{ln}^2)}\,,
}
with ${A = a+s}$.  However, such integrals are clearly singular, generally having a pole of order $r$ at the origin.  It must then be that the singularities cancel in the sum of terms in Eq.~\eqref{eq:Knl-sum}, since $K^n_l$ must be nonsingular (following from the fact that it is proportional to the Radon transform of the bounded function $\Psi^n_{lm}$).

We can therefore subtract the singular contributions from each of the integrals of the form in Eq.~\eqref{eq:singular} to yield nonsingular integrals; substituting these for the singular integrals in Eq.~\eqref{eq:Knl-sum} should then identically yield $K^n_l$, since the singular contributions would have canceled in the sum regardless.  

Thus, we must then calculate nonsingular integrals of the form
\es{eq:nonsingular}{
I(A,r,x_{ln}) =  \mathcal{P} \int\! dx\, \frac{e^{i A x} - \sum_{p=0}^{r-2} \frac{(iAx)^p}{p!}}{x^r (x^2-x_{ln}^2)}\,.
}
The Cauchy principal value of this improper integral can be found using contour integration and the residue theorem.  We let $x$ be a complex number, and then integrate the integrand of Eq.~\eqref{eq:nonsingular} over a contour that includes the real line -- avoiding the simple poles at ${x = 0, \pm x_{ln}}$ -- and closes in either the upper or lower half-plane, depending on the sign of $A$.  Done appropriately, this contour encloses no poles and hence the contour integral vanishes; furthermore, since the integral over the portion of the contour closing in the upper or lower half-plane also vanishes, we have
\begin{widetext}
\es{eq:ReI-ImI}{
I(A,r,x_{ln}) = i \pi \operatorname{sgn} A \sum_{x_i = 0, \pm x_{ln}} \operatorname{Res}\left(\frac{e^{i A x} - \sum_{p=0}^{r-2} \frac{(iAx)^p}{p!}}{x^r (x^2-x_{ln}^2)},\, x_i\right)\,.
}
\end{widetext}

Calculating the residues, we then find
\begin{widetext}
\es{}{
\operatorname{Re} I(A,r,x_{ln}) = \pi \operatorname{sgn} A \left[ \frac{(-1)^{r/2+1} A^{r-1}}{(r-1)!\, x_{ln}^2} - \frac{1}{x_{ln}^{r+1}} \left( \sin Ax_{ln} + \sum_{p=1, \mathrm{odd}}^{r-3} \frac{(-1)^{\frac{p+1}{2}} (Ax_{ln})^p}{p!} \right) \right]\,,\ r\ \mathrm{even}\,\,\\
\operatorname{Im} I(A,r,x_{ln}) = \pi \operatorname{sgn} A \left[ \frac{(-1)^{\frac{r+1}{2}} A^{r-1}}{(r-1)!\, x_{ln}^2} + \frac{1}{x_{ln}^{r+1}} \left( \cos Ax_{ln} + \sum_{p=0, \mathrm{even}}^{r-3} \frac{(-1)^{p/2+1} (Ax_{ln})^p}{p!} \right) \right]\,,\ r\ \mathrm{odd}\,.
}
\end{widetext}
By substituting these functions for the corresponding integrals containing cosine and sine terms in Eq.~\eqref{eq:Knl-sum}, we arrive at the final expression for $K^n_l$ given in Eq.~\eqref{eq:Knl}.  As advertised, we see that $K^n_l$ is a finite sum of elementary functions.

Furthermore, as previously mentioned, we see that $i^{-l} K^n_l$ is always real.  Interestingly enough, this results because when $l$ is even, the coefficients $\alpha_{lr}$ and $\beta_{lr}$ vanish for odd and even values of $r$, respectively, and vice versa when $l$ is odd.  Since $\operatorname{Re} I(A,r,x_{ln})$ and $\operatorname{Im} I(A,r,x_{ln})$ vanish for odd and even $r$, respectively, we see that the sum in Eq.~\eqref{eq:Knl-sum} fixes $K^n_l$ to be real for even $l$ and imaginary for odd $l$, so that $i^{-l} K^n_l$ is indeed always real.

\section{Inverting the Radon Transform} \label{app:invert}

The Radon transform can be inverted by means of the Laplacian operator,
\es{eq:Laplace-inversion}{
f(\mathbf{v}) &= -\frac{1}{8\pi^2} \nabla^2_{\mathbf{v}} \int\! \dOq\, \widehat{f}(\mathbf{v}\cdot\qhat, \qhat)\\
&= -\frac{1}{8\pi^2} \int\! \dOq\, \widehat{f}''(\mathbf{v}\cdot\qhat, \qhat)\,,
}
where ${\widehat{f}''(\vq, \uhat) = d^2\widehat{f}(\vq,\uhat)/d\vq^2}$ \cite{hep-ph/0209110,deans2007radon}.  This relation implies that we can recover $f$ at a given $\mathbf{v}$ if we have a sufficient number of events at ${\vq \leq v}$.  Unfortunately, standard tomographic algorithms that make use of this relation will be difficult to apply to directional experiments, for two reasons: 1) the lack of perfect energy resolution precludes the complete detection of events down to ${\vq = 0}$, and 2) the lack of nearly continuous data will make difficult the evaluation of the second derivative $\widehat{f}''$.

Nevertheless, such inversion formulas have been studied in the context of directional experiments.  For example, consider the result found in Ref.~\cite{hep-ph/0209110},
\es{eq:Gondolo-inversion}{
f_{lm}(v) = -\frac{1}{2\pi v} \int_0^v\!\! d\vq\, P_l\left(\frac{\vq}{v}\right) \widehat{f}''_{lm}(\vq)\,,
} 
where $f_{lm}(v)$ is a speed-dependent spherical-harmonic coefficient for the lab-frame velocity distribution $f$ and $P_l(x)$ is a Legendre polynomial.  The corresponding spherical-harmonic coefficients $g_{lm}(v)$ for the Galactic-frame velocity distribution can then be calculated by performing the translation by $\mathbf{\vlab}$ appropriately.  Eq.~\eqref{eq:Gondolo-inversion} then relates the coefficients $g_{lm}(v)$ to the observable $\widehat{f}$.  In practice, the true observable would be the directional recoil spectrum $dR/dE\dOq$; however, if the WIMP mass and the nuclear form factor are known, it is straightforward to use Eq.~\eqref{eq:spectrum} in this expression to replace $\widehat{f}_{lm}(\vq)$ in favor of $\left(dR/dE\dOq\right)_{lm}$.

We can investigate whether the Fourier-Bessel basis yields a similar inversion formula for the Fourier-Bessel coefficients $g^n_{lm}$ in terms of $\widehat{f}$.  To do so, we shall again exploit the relation of the Fourier-Bessel basis to the Fourier basis and the plane-wave expansion.  Since $f$ and $g$ are related by a translation in velocity space as in Eq.~\eqref{eq:f-g}, their respective Fourier coefficients are related by a phase shift,
\es{}{
\widetilde{f}(\mathbf{u}) = e^{i \mathbf{u}\cdot\mathbf{\vlab}} \widetilde{g}(\mathbf{u})\,.
}
We may then use the Fourier slice theorem of Eq.~\eqref{eq:Fourier-slice} to write the Fourier coefficients for $g$ as
\es{}{
\widetilde{g}(\mathbf{u}) = e^{-i \mathbf{u}\cdot\mathbf{\vlab}} \int\! d\vq\, e^{-iu\vq} \widehat{f}(\vq, \uhat) \,.
}
Finally, we use the relation between the Fourier and Fourier-Bessel coefficients in Eq.~\eqref{eq:FB-F-coeff} to find
\es{eq:FB-F-FST}{
g^u_{lm} &= i^l \int\! \dOu\, S_{lm}(\uhat)\, e^{-i \mathbf{u}\cdot\mathbf{\vlab}}\\
&\indent\indent\times \int\! d\vq\, e^{-iu\vq} \widehat{f}(\vq, \uhat) \,.
}

We have thus found a relation between $g^u_{lm}$ and $\widehat{f}$, which we can use with Eq.~\eqref{eq:spectrum} to write the former in terms of the observable $dR/dE\dOq$.  However, we can massage this relation further by using the complex conjugate of Eq.~\eqref{eq:pwe} to expand $e^{-i\mathbf{u}\cdot\mathbf{\vlab}}$ (choosing $l'$ and $m'$ as the indices of summation) and decomposing the angular dependence of $\widehat{f}(\vq, \uhat)$ using real spherical harmonics,
\es{eq:fhat-exp}{
\widehat{f}(\vq, \uhat) = \sum_{l''m''} \widehat{f}_{l''m''}(\vq) S_{l''m''}(\uhat)\,. 
}
Using also the relation in Eq.~\eqref{eq:discCoeff}, this procedure gives
\begin{widetext}
\es{eq:FB-coeff-result}{
g^n_{lm} &= i^l c_{ln} \sum_{\substack{{l'm'}\\{l''m''}}} i^{-l'} \left[\int\! \dOu\, S_{lm}(\uhat) S_{l'm'}(\uhat) S_{l''m''}(\uhat) \right] \Psi^u_{l'm'}(\mathbf{\vlab}) \int\! d\vq\, e^{-iu_{ln}\vq} \widehat{f}_{l''m''}(\vq)\\
	&= i^l c_{ln} \sum_{\substack{{l'm'}\\{l''m''}}} i^{-l'} H(L L' L'') \Psi^u_{l'm'}(\mathbf{\vlab}) \int\! d\vq\, e^{-iu_{ln}\vq} \widehat{f}_{l''m''}(\vq)\,. 
}
\end{widetext}
Again, see Appendix~\ref{app:realSH} for the definition of the Gaunt-like coefficients $H(L L' L'')$.

Eq.~\eqref{eq:FB-coeff-result} gives an expression for the Fourier-Bessel coefficients $g^n_{lm}$ in terms of products of the basis functions (evaluated at the Galactic-frame lab velocity $\mathbf{\vlab}$) and Fourier transforms of the real-spherical-harmonic coefficients $\widehat{f}_{l''m''}(\vq)$.  Examining the case ${\mathbf{\vlab} = 0}$, it becomes clear that Eq.~\eqref{eq:FB-coeff-result} is simply a restatement of the Fourier slice theorem.

We have thus found an inversion formula for the Fourier-Bessel coefficients $g^n_{lm}$.  Comparing Eq.~\eqref{eq:Gondolo-inversion} with Eq.~\eqref{eq:FB-coeff-result}, we see that both inversion formulas involve an integral of a function of the observable $\widehat{f}_{lm}(\vq)$ over $\vq$.  Further inspection of Eq.~\eqref{eq:Gondolo-inversion} shows that the integral involved there contains $\widehat{f}''_{lm}(\vq)$, in contrast with the integral containing $\widehat{f}_{lm}(\vq)$ found in our inversion formula.  Thus, by instead calculating the Fourier-Bessel coefficients using Eq.~\eqref{eq:FB-coeff-result}, we avoid the second-derivative term that arises when calculating the spherical-harmonic coefficients using Eq.~\eqref{eq:Gondolo-inversion}.  This is clearly preferable if the number of events is not large, in which case determination of the second derivative of the observable may be difficult.

However, in both formulas the integral extends to ${\vq = 0}$, implying that perfect energy resolution is required to accurately recover the velocity-distribution expansion coefficients.  Furthermore, the integral present in our inversion formula extends to all $\vq$, requiring spectral coverage at higher energies where the signal may be obscured by unavoidable backgrounds.  This is simply a side effect of attempting to characterize the velocity dependence of the distribution via the expansion coefficients; i.e., since the Fourier-Bessel coefficients are determined by an integral over all velocities $\mathbf{v}$ as in Eq.~\eqref{eq:discCoeff}, it follows that one requires data at all $\vq$ in order to recover them.

It is clear that direct inversion of the Radon transform to recover the velocity distribution from the directional recoil spectrum is problematic in several respects.  Using a maximum-likelihood method that scans over the parameter space of expansion coefficients may then be the best option available for reconstructing the velocity distribution from a limited number of events.

\bibliographystyle{apsrev}
\bibliography{directional-tam}

\begin{thebibliography}{46}
\expandafter\ifx\csname natexlab\endcsname\relax\def\natexlab#1{#1}\fi
\expandafter\ifx\csname bibnamefont\endcsname\relax
  \def\bibnamefont#1{#1}\fi
\expandafter\ifx\csname bibfnamefont\endcsname\relax
  \def\bibfnamefont#1{#1}\fi
\expandafter\ifx\csname citenamefont\endcsname\relax
  \def\citenamefont#1{#1}\fi
\expandafter\ifx\csname url\endcsname\relax
  \def\url#1{\texttt{#1}}\fi
\expandafter\ifx\csname urlprefix\endcsname\relax\def\urlprefix{URL }\fi
\providecommand{\bibinfo}[2]{#2}
\providecommand{\eprint}[2][]{\url{#2}}

\bibitem[{\citenamefont{Ahlen et~al.}(2010)\citenamefont{Ahlen, Afshordi,
  Battat, Billard, Bozorgnia et~al.}}]{arXiv:0911.0323}
\bibinfo{author}{\bibfnamefont{S.}~\bibnamefont{Ahlen}},
  \bibinfo{author}{\bibfnamefont{N.}~\bibnamefont{Afshordi}},
  \bibinfo{author}{\bibfnamefont{J.}~\bibnamefont{Battat}},
  \bibinfo{author}{\bibfnamefont{J.}~\bibnamefont{Billard}},
  \bibinfo{author}{\bibfnamefont{N.}~\bibnamefont{Bozorgnia}},
  \bibnamefont{et~al.}, \bibinfo{journal}{Int.J.Mod.Phys.}
  \textbf{\bibinfo{volume}{A25}}, \bibinfo{pages}{1} (\bibinfo{year}{2010}),
  \eprint{0911.0323}.

\bibitem[{\citenamefont{Spergel}(1988)}]{Spergel:1987kx}
\bibinfo{author}{\bibfnamefont{D.~N.} \bibnamefont{Spergel}},
  \bibinfo{journal}{Phys.Rev.} \textbf{\bibinfo{volume}{D37}},
  \bibinfo{pages}{1353} (\bibinfo{year}{1988}).

\bibitem[{\citenamefont{Drukier et~al.}(1986)\citenamefont{Drukier, Freese, and
  Spergel}}]{Drukier:1986tm}
\bibinfo{author}{\bibfnamefont{A.}~\bibnamefont{Drukier}},
  \bibinfo{author}{\bibfnamefont{K.}~\bibnamefont{Freese}}, \bibnamefont{and}
  \bibinfo{author}{\bibfnamefont{D.}~\bibnamefont{Spergel}},
  \bibinfo{journal}{Phys.Rev.} \textbf{\bibinfo{volume}{D33}},
  \bibinfo{pages}{3495} (\bibinfo{year}{1986}).

\bibitem[{\citenamefont{Gondolo}(2002)}]{hep-ph/0209110}
\bibinfo{author}{\bibfnamefont{P.}~\bibnamefont{Gondolo}},
  \bibinfo{journal}{Phys.Rev.} \textbf{\bibinfo{volume}{D66}},
  \bibinfo{pages}{103513} (\bibinfo{year}{2002}), \eprint{hep-ph/0209110}.

\bibitem[{\citenamefont{Alenazi and Gondolo}(2008)}]{arXiv:0712.0053}
\bibinfo{author}{\bibfnamefont{M.~S.} \bibnamefont{Alenazi}} \bibnamefont{and}
  \bibinfo{author}{\bibfnamefont{P.}~\bibnamefont{Gondolo}},
  \bibinfo{journal}{Phys.Rev.} \textbf{\bibinfo{volume}{D77}},
  \bibinfo{pages}{043532} (\bibinfo{year}{2008}), \eprint{0712.0053}.

\bibitem[{\citenamefont{Billard et~al.}(2012)\citenamefont{Billard, Mayet, and
  Santos}}]{arXiv:1110.0951}
\bibinfo{author}{\bibfnamefont{J.}~\bibnamefont{Billard}},
  \bibinfo{author}{\bibfnamefont{F.}~\bibnamefont{Mayet}}, \bibnamefont{and}
  \bibinfo{author}{\bibfnamefont{D.}~\bibnamefont{Santos}},
  \bibinfo{journal}{EAS Publ.Ser.} \textbf{\bibinfo{volume}{53}},
  \bibinfo{pages}{67} (\bibinfo{year}{2012}), \eprint{1110.0951}.

\bibitem[{\citenamefont{Lee and Peter}(2012)}]{arXiv:1202.5035}
\bibinfo{author}{\bibfnamefont{S.~K.} \bibnamefont{Lee}} \bibnamefont{and}
  \bibinfo{author}{\bibfnamefont{A.~H.} \bibnamefont{Peter}},
  \bibinfo{journal}{JCAP} \textbf{\bibinfo{volume}{1204}}, \bibinfo{pages}{029}
  (\bibinfo{year}{2012}), \eprint{1202.5035}.

\bibitem[{\citenamefont{Peter}(2011)}]{arXiv:1103.5145}
\bibinfo{author}{\bibfnamefont{A.~H.} \bibnamefont{Peter}},
  \bibinfo{journal}{Phys.Rev.} \textbf{\bibinfo{volume}{D83}},
  \bibinfo{pages}{125029} (\bibinfo{year}{2011}), \eprint{1103.5145}.

\bibitem[{\citenamefont{Kavanagh and Green}(2012)}]{arXiv:1207.2039}
\bibinfo{author}{\bibfnamefont{B.~J.} \bibnamefont{Kavanagh}} \bibnamefont{and}
  \bibinfo{author}{\bibfnamefont{A.~M.} \bibnamefont{Green}},
  \bibinfo{journal}{Phys.Rev.} \textbf{\bibinfo{volume}{D86}},
  \bibinfo{pages}{065027} (\bibinfo{year}{2012}), \eprint{1207.2039}.

\bibitem[{\citenamefont{Kavanagh and Green}(2013)}]{arXiv:1303.6868}
\bibinfo{author}{\bibfnamefont{B.~J.} \bibnamefont{Kavanagh}} \bibnamefont{and}
  \bibinfo{author}{\bibfnamefont{A.~M.} \bibnamefont{Green}},
  \bibinfo{journal}{Phys.Rev.Lett.} \textbf{\bibinfo{volume}{111}},
  \bibinfo{pages}{031302} (\bibinfo{year}{2013}), \eprint{1303.6868}.

\bibitem[{\citenamefont{Peter et~al.}(2013)\citenamefont{Peter, Gluscevic,
  Green, Kavanagh, and Lee}}]{arXiv:1310.7039}
\bibinfo{author}{\bibfnamefont{A.~H.~G.} \bibnamefont{Peter}},
  \bibinfo{author}{\bibfnamefont{V.}~\bibnamefont{Gluscevic}},
  \bibinfo{author}{\bibfnamefont{A.~M.} \bibnamefont{Green}},
  \bibinfo{author}{\bibfnamefont{B.~J.} \bibnamefont{Kavanagh}},
  \bibnamefont{and} \bibinfo{author}{\bibfnamefont{S.~K.} \bibnamefont{Lee}}
  (\bibinfo{year}{2013}), \eprint{1310.7039}.

\bibitem[{\citenamefont{Kavanagh}(2013)}]{arXiv:1312.1852}
\bibinfo{author}{\bibfnamefont{B.~J.} \bibnamefont{Kavanagh}}
  (\bibinfo{year}{2013}), \eprint{1312.1852}.

\bibitem[{\citenamefont{Alves et~al.}(2012)\citenamefont{Alves, Hedri, and
  Wacker}}]{arXiv:1204.5487}
\bibinfo{author}{\bibfnamefont{D.~S.} \bibnamefont{Alves}},
  \bibinfo{author}{\bibfnamefont{S.~E.} \bibnamefont{Hedri}}, \bibnamefont{and}
  \bibinfo{author}{\bibfnamefont{J.~G.} \bibnamefont{Wacker}}
  (\bibinfo{year}{2012}), \eprint{1204.5487}.

\bibitem[{\citenamefont{Deans}(2007)}]{deans2007radon}
\bibinfo{author}{\bibfnamefont{S.}~\bibnamefont{Deans}},
  \emph{\bibinfo{title}{The Radon Transform and Some of Its Applications}}
  (\bibinfo{publisher}{Dover Publications}, \bibinfo{year}{2007}).

\bibitem[{\citenamefont{Lisanti et~al.}(2011)\citenamefont{Lisanti, Strigari,
  Wacker, and Wechsler}}]{arXiv:1010.4300}
\bibinfo{author}{\bibfnamefont{M.}~\bibnamefont{Lisanti}},
  \bibinfo{author}{\bibfnamefont{L.~E.} \bibnamefont{Strigari}},
  \bibinfo{author}{\bibfnamefont{J.~G.} \bibnamefont{Wacker}},
  \bibnamefont{and} \bibinfo{author}{\bibfnamefont{R.~H.}
  \bibnamefont{Wechsler}}, \bibinfo{journal}{Phys.Rev.}
  \textbf{\bibinfo{volume}{D83}}, \bibinfo{pages}{023519}
  (\bibinfo{year}{2011}), \eprint{1010.4300}.

\bibitem[{\citenamefont{Mao et~al.}(2013)\citenamefont{Mao, Strigari, and
  Wechsler}}]{arXiv:1304.6401}
\bibinfo{author}{\bibfnamefont{Y.-Y.} \bibnamefont{Mao}},
  \bibinfo{author}{\bibfnamefont{L.~E.} \bibnamefont{Strigari}},
  \bibnamefont{and} \bibinfo{author}{\bibfnamefont{R.~H.}
  \bibnamefont{Wechsler}} (\bibinfo{year}{2013}), \eprint{1304.6401}.

\bibitem[{\citenamefont{Green}(2002)}]{astro-ph/0209528}
\bibinfo{author}{\bibfnamefont{A.~M.} \bibnamefont{Green}}, pp.
  \bibinfo{pages}{78--83} (\bibinfo{year}{2002}), \eprint{astro-ph/0209528}.

\bibitem[{\citenamefont{Host and Hansen}(2007)}]{arXiv:0704.2909}
\bibinfo{author}{\bibfnamefont{O.}~\bibnamefont{Host}} \bibnamefont{and}
  \bibinfo{author}{\bibfnamefont{S.~H.} \bibnamefont{Hansen}},
  \bibinfo{journal}{JCAP} \textbf{\bibinfo{volume}{0706}}, \bibinfo{pages}{016}
  (\bibinfo{year}{2007}), \eprint{0704.2909}.

\bibitem[{\citenamefont{Vogelsberger et~al.}(2009)\citenamefont{Vogelsberger,
  Helmi, Springel, White, Wang et~al.}}]{arXiv:0812.0362}
\bibinfo{author}{\bibfnamefont{M.}~\bibnamefont{Vogelsberger}},
  \bibinfo{author}{\bibfnamefont{A.}~\bibnamefont{Helmi}},
  \bibinfo{author}{\bibfnamefont{V.}~\bibnamefont{Springel}},
  \bibinfo{author}{\bibfnamefont{S.~D.} \bibnamefont{White}},
  \bibinfo{author}{\bibfnamefont{J.}~\bibnamefont{Wang}}, \bibnamefont{et~al.},
  \bibinfo{journal}{Mon.Not.Roy.Astron.Soc.} \textbf{\bibinfo{volume}{395}},
  \bibinfo{pages}{797} (\bibinfo{year}{2009}), \eprint{0812.0362}.

\bibitem[{\citenamefont{Fornasa and Green}(2013)}]{arXiv:1311.5477}
\bibinfo{author}{\bibfnamefont{M.}~\bibnamefont{Fornasa}} \bibnamefont{and}
  \bibinfo{author}{\bibfnamefont{A.~M.} \bibnamefont{Green}}
  (\bibinfo{year}{2013}), \eprint{1311.5477}.

\bibitem[{\citenamefont{Read et~al.}(2008)\citenamefont{Read, Lake, Agertz, and
  Debattista}}]{arXiv:0803.2714}
\bibinfo{author}{\bibfnamefont{J.}~\bibnamefont{Read}},
  \bibinfo{author}{\bibfnamefont{G.}~\bibnamefont{Lake}},
  \bibinfo{author}{\bibfnamefont{O.}~\bibnamefont{Agertz}}, \bibnamefont{and}
  \bibinfo{author}{\bibfnamefont{V.~P.} \bibnamefont{Debattista}},
  \bibinfo{journal}{Mon.Not.Roy.Astron.Soc.} \textbf{\bibinfo{volume}{389}},
  \bibinfo{pages}{1041} (\bibinfo{year}{2008}), \eprint{0803.2714}.

\bibitem[{\citenamefont{Bruch et~al.}(2009)\citenamefont{Bruch, Read, Baudis,
  and Lake}}]{arXiv:0804.2896}
\bibinfo{author}{\bibfnamefont{T.}~\bibnamefont{Bruch}},
  \bibinfo{author}{\bibfnamefont{J.}~\bibnamefont{Read}},
  \bibinfo{author}{\bibfnamefont{L.}~\bibnamefont{Baudis}}, \bibnamefont{and}
  \bibinfo{author}{\bibfnamefont{G.}~\bibnamefont{Lake}},
  \bibinfo{journal}{Astrophys.J.} \textbf{\bibinfo{volume}{696}},
  \bibinfo{pages}{920} (\bibinfo{year}{2009}), \eprint{0804.2896}.

\bibitem[{\citenamefont{Read et~al.}(2009)\citenamefont{Read, Mayer, Brooks,
  Governato, and Lake}}]{arXiv:0902.0009}
\bibinfo{author}{\bibfnamefont{J.}~\bibnamefont{Read}},
  \bibinfo{author}{\bibfnamefont{L.}~\bibnamefont{Mayer}},
  \bibinfo{author}{\bibfnamefont{A.}~\bibnamefont{Brooks}},
  \bibinfo{author}{\bibfnamefont{F.}~\bibnamefont{Governato}},
  \bibnamefont{and} \bibinfo{author}{\bibfnamefont{G.}~\bibnamefont{Lake}}
  (\bibinfo{year}{2009}), \eprint{0902.0009}.

\bibitem[{\citenamefont{Billard et~al.}(2013)\citenamefont{Billard, Riffard,
  Mayet, and Santos}}]{arXiv:1207.1050}
\bibinfo{author}{\bibfnamefont{J.}~\bibnamefont{Billard}},
  \bibinfo{author}{\bibfnamefont{Q.}~\bibnamefont{Riffard}},
  \bibinfo{author}{\bibfnamefont{F.}~\bibnamefont{Mayet}}, \bibnamefont{and}
  \bibinfo{author}{\bibfnamefont{D.}~\bibnamefont{Santos}},
  \bibinfo{journal}{Phys.Lett.} \textbf{\bibinfo{volume}{B718}},
  \bibinfo{pages}{1171} (\bibinfo{year}{2013}), \eprint{1207.1050}.

\bibitem[{\citenamefont{Fan et~al.}(2013{\natexlab{a}})\citenamefont{Fan, Katz,
  Randall, and Reece}}]{arXiv:1303.1521}
\bibinfo{author}{\bibfnamefont{J.}~\bibnamefont{Fan}},
  \bibinfo{author}{\bibfnamefont{A.}~\bibnamefont{Katz}},
  \bibinfo{author}{\bibfnamefont{L.}~\bibnamefont{Randall}}, \bibnamefont{and}
  \bibinfo{author}{\bibfnamefont{M.}~\bibnamefont{Reece}},
  \bibinfo{journal}{Phys.Dark Univ.} \textbf{\bibinfo{volume}{2}},
  \bibinfo{pages}{139} (\bibinfo{year}{2013}{\natexlab{a}}),
  \eprint{1303.1521}.

\bibitem[{\citenamefont{Fan et~al.}(2013{\natexlab{b}})\citenamefont{Fan, Katz,
  Randall, and Reece}}]{arXiv:1303.3271}
\bibinfo{author}{\bibfnamefont{J.}~\bibnamefont{Fan}},
  \bibinfo{author}{\bibfnamefont{A.}~\bibnamefont{Katz}},
  \bibinfo{author}{\bibfnamefont{L.}~\bibnamefont{Randall}}, \bibnamefont{and}
  \bibinfo{author}{\bibfnamefont{M.}~\bibnamefont{Reece}},
  \bibinfo{journal}{Phys.Rev.Lett.} \textbf{\bibinfo{volume}{110}},
  \bibinfo{pages}{211302} (\bibinfo{year}{2013}{\natexlab{b}}),
  \eprint{1303.3271}.

\bibitem[{\citenamefont{McCullough and Randall}(2013)}]{arXiv:1307.4095}
\bibinfo{author}{\bibfnamefont{M.}~\bibnamefont{McCullough}} \bibnamefont{and}
  \bibinfo{author}{\bibfnamefont{L.}~\bibnamefont{Randall}},
  \bibinfo{journal}{JCAP} \textbf{\bibinfo{volume}{1310}}, \bibinfo{pages}{058}
  (\bibinfo{year}{2013}), \eprint{1307.4095}.

\bibitem[{\citenamefont{Kuhlen et~al.}(2013)\citenamefont{Kuhlen, Pillepich,
  Guedes, and Madau}}]{arXiv:1308.1703}
\bibinfo{author}{\bibfnamefont{M.}~\bibnamefont{Kuhlen}},
  \bibinfo{author}{\bibfnamefont{A.}~\bibnamefont{Pillepich}},
  \bibinfo{author}{\bibfnamefont{J.}~\bibnamefont{Guedes}}, \bibnamefont{and}
  \bibinfo{author}{\bibfnamefont{P.}~\bibnamefont{Madau}}
  (\bibinfo{year}{2013}), \eprint{1308.1703}.

\bibitem[{\citenamefont{Freese et~al.}(2005)\citenamefont{Freese, Gondolo, and
  Newberg}}]{astro-ph/0309279}
\bibinfo{author}{\bibfnamefont{K.}~\bibnamefont{Freese}},
  \bibinfo{author}{\bibfnamefont{P.}~\bibnamefont{Gondolo}}, \bibnamefont{and}
  \bibinfo{author}{\bibfnamefont{H.~J.} \bibnamefont{Newberg}},
  \bibinfo{journal}{Phys.Rev.} \textbf{\bibinfo{volume}{D71}},
  \bibinfo{pages}{043516} (\bibinfo{year}{2005}), \eprint{astro-ph/0309279}.

\bibitem[{\citenamefont{Savage et~al.}(2006)\citenamefont{Savage, Freese, and
  Gondolo}}]{astro-ph/0607121}
\bibinfo{author}{\bibfnamefont{C.}~\bibnamefont{Savage}},
  \bibinfo{author}{\bibfnamefont{K.}~\bibnamefont{Freese}}, \bibnamefont{and}
  \bibinfo{author}{\bibfnamefont{P.}~\bibnamefont{Gondolo}},
  \bibinfo{journal}{Phys.Rev.} \textbf{\bibinfo{volume}{D74}},
  \bibinfo{pages}{043531} (\bibinfo{year}{2006}), \eprint{astro-ph/0607121}.

\bibitem[{\citenamefont{Natarajan et~al.}(2011)\citenamefont{Natarajan, Savage,
  and Freese}}]{arXiv:1109.0014}
\bibinfo{author}{\bibfnamefont{A.}~\bibnamefont{Natarajan}},
  \bibinfo{author}{\bibfnamefont{C.}~\bibnamefont{Savage}}, \bibnamefont{and}
  \bibinfo{author}{\bibfnamefont{K.}~\bibnamefont{Freese}},
  \bibinfo{journal}{Phys.Rev.} \textbf{\bibinfo{volume}{D84}},
  \bibinfo{pages}{103005} (\bibinfo{year}{2011}), \eprint{1109.0014}.

\bibitem[{\citenamefont{Purcell et~al.}(2012)\citenamefont{Purcell, Zentner,
  and Wang}}]{arXiv:1203.6617}
\bibinfo{author}{\bibfnamefont{C.~W.} \bibnamefont{Purcell}},
  \bibinfo{author}{\bibfnamefont{A.~R.} \bibnamefont{Zentner}},
  \bibnamefont{and} \bibinfo{author}{\bibfnamefont{M.-Y.} \bibnamefont{Wang}},
  \bibinfo{journal}{JCAP} \textbf{\bibinfo{volume}{1208}}, \bibinfo{pages}{027}
  (\bibinfo{year}{2012}), \eprint{1203.6617}.

\bibitem[{\citenamefont{Lisanti and Spergel}(2012)}]{arXiv:1105.4166}
\bibinfo{author}{\bibfnamefont{M.}~\bibnamefont{Lisanti}} \bibnamefont{and}
  \bibinfo{author}{\bibfnamefont{D.~N.} \bibnamefont{Spergel}},
  \bibinfo{journal}{Phys.Dark Univ.} \textbf{\bibinfo{volume}{1}},
  \bibinfo{pages}{155} (\bibinfo{year}{2012}), \eprint{1105.4166}.

\bibitem[{\citenamefont{Kuhlen et~al.}(2012)\citenamefont{Kuhlen, Lisanti, and
  Spergel}}]{arXiv:1202.0007}
\bibinfo{author}{\bibfnamefont{M.}~\bibnamefont{Kuhlen}},
  \bibinfo{author}{\bibfnamefont{M.}~\bibnamefont{Lisanti}}, \bibnamefont{and}
  \bibinfo{author}{\bibfnamefont{D.~N.} \bibnamefont{Spergel}},
  \bibinfo{journal}{Phys.Rev.} \textbf{\bibinfo{volume}{D86}},
  \bibinfo{pages}{063505} (\bibinfo{year}{2012}), \eprint{1202.0007}.

\bibitem[{\citenamefont{Freese et~al.}(2001)\citenamefont{Freese, Gondolo, and
  Stodolsky}}]{astro-ph/0106480}
\bibinfo{author}{\bibfnamefont{K.}~\bibnamefont{Freese}},
  \bibinfo{author}{\bibfnamefont{P.}~\bibnamefont{Gondolo}}, \bibnamefont{and}
  \bibinfo{author}{\bibfnamefont{L.}~\bibnamefont{Stodolsky}},
  \bibinfo{journal}{Phys.Rev.} \textbf{\bibinfo{volume}{D64}},
  \bibinfo{pages}{123502} (\bibinfo{year}{2001}), \eprint{astro-ph/0106480}.

\bibitem[{\citenamefont{Baushev}(2013)}]{arXiv:1208.0392}
\bibinfo{author}{\bibfnamefont{A.}~\bibnamefont{Baushev}},
  \bibinfo{journal}{Astrophys.J.} \textbf{\bibinfo{volume}{771}},
  \bibinfo{pages}{117} (\bibinfo{year}{2013}), \eprint{1208.0392}.

\bibitem[{\citenamefont{Fisher et~al.}(1994)\citenamefont{Fisher, Lahav,
  Hoffman, Lynden-Bell, and Zaroubi}}]{astro-ph/9406009}
\bibinfo{author}{\bibfnamefont{K.}~\bibnamefont{Fisher}},
  \bibinfo{author}{\bibfnamefont{O.}~\bibnamefont{Lahav}},
  \bibinfo{author}{\bibfnamefont{Y.}~\bibnamefont{Hoffman}},
  \bibinfo{author}{\bibfnamefont{D.}~\bibnamefont{Lynden-Bell}},
  \bibnamefont{and} \bibinfo{author}{\bibfnamefont{S.}~\bibnamefont{Zaroubi}}
  (\bibinfo{year}{1994}), \eprint{astro-ph/9406009}.

\bibitem[{\citenamefont{Leistedt et~al.}(2011)\citenamefont{Leistedt, Rassat,
  Refregier, and Starck}}]{arXiv:1111.3591}
\bibinfo{author}{\bibfnamefont{B.}~\bibnamefont{Leistedt}},
  \bibinfo{author}{\bibfnamefont{A.}~\bibnamefont{Rassat}},
  \bibinfo{author}{\bibfnamefont{A.}~\bibnamefont{Refregier}},
  \bibnamefont{and} \bibinfo{author}{\bibfnamefont{J.}~\bibnamefont{Starck}}
  (\bibinfo{year}{2011}), \eprint{1111.3591}.

\bibitem[{\citenamefont{Lanusse et~al.}(2011)\citenamefont{Lanusse, Rassat, and
  Starck}}]{arXiv:1112.0561}
\bibinfo{author}{\bibfnamefont{F.}~\bibnamefont{Lanusse}},
  \bibinfo{author}{\bibfnamefont{A.}~\bibnamefont{Rassat}}, \bibnamefont{and}
  \bibinfo{author}{\bibfnamefont{J.}~\bibnamefont{Starck}}
  (\bibinfo{year}{2011}), \eprint{1112.0561}.

\bibitem[{\citenamefont{Leistedt and McEwen}(2012)}]{arXiv:1205.0792}
\bibinfo{author}{\bibfnamefont{B.}~\bibnamefont{Leistedt}} \bibnamefont{and}
  \bibinfo{author}{\bibfnamefont{J.}~\bibnamefont{McEwen}},
  \bibinfo{journal}{IEEE Trans. Signal. Process. ,}
  \textbf{\bibinfo{volume}{60}}, \bibinfo{pages}{, 6257}
  (\bibinfo{year}{2012}), \eprint{1205.0792}.

\bibitem[{\citenamefont{Leistedt et~al.}(2013)\citenamefont{Leistedt, Peiris,
  and McEwen}}]{arXiv:1308.5480}
\bibinfo{author}{\bibfnamefont{B.}~\bibnamefont{Leistedt}},
  \bibinfo{author}{\bibfnamefont{H.~V.} \bibnamefont{Peiris}},
  \bibnamefont{and} \bibinfo{author}{\bibfnamefont{J.~D.} \bibnamefont{McEwen}}
  (\bibinfo{year}{2013}), \eprint{1308.5480}.

\bibitem[{\citenamefont{Chang et~al.}(2012)\citenamefont{Chang, Pradler, and
  Yavin}}]{arXiv:1111.4222}
\bibinfo{author}{\bibfnamefont{S.}~\bibnamefont{Chang}},
  \bibinfo{author}{\bibfnamefont{J.}~\bibnamefont{Pradler}}, \bibnamefont{and}
  \bibinfo{author}{\bibfnamefont{I.}~\bibnamefont{Yavin}},
  \bibinfo{journal}{Phys.Rev.} \textbf{\bibinfo{volume}{D85}},
  \bibinfo{pages}{063505} (\bibinfo{year}{2012}), \eprint{1111.4222}.

\bibitem[{\citenamefont{Lee et~al.}(2013)\citenamefont{Lee, Lisanti, and
  Safdi}}]{arXiv:1307.5323}
\bibinfo{author}{\bibfnamefont{S.~K.} \bibnamefont{Lee}},
  \bibinfo{author}{\bibfnamefont{M.}~\bibnamefont{Lisanti}}, \bibnamefont{and}
  \bibinfo{author}{\bibfnamefont{B.~R.} \bibnamefont{Safdi}},
  \bibinfo{journal}{JCAP} \textbf{\bibinfo{volume}{1311}}, \bibinfo{pages}{033}
  (\bibinfo{year}{2013}), \eprint{1307.5323}.

\bibitem[{\citenamefont{Lee et~al.}(2014)\citenamefont{Lee, Lisanti, Peter, and
  Safdi}}]{arXiv:1308.1953}
\bibinfo{author}{\bibfnamefont{S.~K.} \bibnamefont{Lee}},
  \bibinfo{author}{\bibfnamefont{M.}~\bibnamefont{Lisanti}},
  \bibinfo{author}{\bibfnamefont{A.~H.~G.} \bibnamefont{Peter}},
  \bibnamefont{and} \bibinfo{author}{\bibfnamefont{B.~R.} \bibnamefont{Safdi}},
  \bibinfo{journal}{Phys.Rev.Lett.} \textbf{\bibinfo{volume}{112}},
  \bibinfo{pages}{011301} (\bibinfo{year}{2014}), \eprint{1308.1953}.

\bibitem[{\citenamefont{Dai et~al.}(2012)\citenamefont{Dai, Kamionkowski, and
  Jeong}}]{arXiv:1209.0761}
\bibinfo{author}{\bibfnamefont{L.}~\bibnamefont{Dai}},
  \bibinfo{author}{\bibfnamefont{M.}~\bibnamefont{Kamionkowski}},
  \bibnamefont{and} \bibinfo{author}{\bibfnamefont{D.}~\bibnamefont{Jeong}},
  \bibinfo{journal}{Phys.Rev.} \textbf{\bibinfo{volume}{D86}},
  \bibinfo{pages}{125013} (\bibinfo{year}{2012}), \eprint{1209.0761}.

\bibitem[{\citenamefont{Haynes and Payne}(1997)}]{Haynes1997}
\bibinfo{author}{\bibfnamefont{P.~D.} \bibnamefont{Haynes}} \bibnamefont{and}
  \bibinfo{author}{\bibfnamefont{M.~C.} \bibnamefont{Payne}},
  \bibinfo{journal}{{Computer Physics Communications}}
  \textbf{\bibinfo{volume}{102}}, \bibinfo{pages}{17} (\bibinfo{year}{1997}).

\end{thebibliography}

\end{document}